\documentclass{article}

\usepackage{graphicx}

\usepackage[latin1]{inputenc}
\usepackage{amssymb}

\renewcommand{\vec}[1]{{\mathbf #1}}

\begin{document}

\begin{center}

\vspace{1cm}

\LARGE
Topology effects on the heat capacity of mesoscopic superconducting disks\\

\vspace{2cm}

\large
F.R. Ong\footnote{florian.ong@grenoble.cnrs.fr} and O. Bourgeois\footnote{olivier.bourgeois@grenoble.cnrs.fr}\\

\vspace{1cm}

\normalsize
\it
Institut Néel, CNRS-UJF, 25 av des Martyrs BP166, 38042 Grenoble Cedex 9, France

\end{center}

\vspace{2cm}

\abstract{
Phase transitions in superconducting mesoscopic disks have been studied over the $H-T$ phase diagram through heat capacity measurement of an array of independent aluminium disks. These disks exhibit non periodic modulations versus $H$ of the height of the heat capacity jump at the superconducting to normal transition. This behaviour is attributed to giant vortex states characterized by their vorticity $L$. A crossover from a bulk-like to a mesoscopic behaviour is demonstrated. $C_{\rm p}$ versus $H$ plots exhibit cascades of phase transitions as $L$ increases or decreases by one unity, with a strong hysteresis. Phase diagrams of giant vortex states inside the superconducting region are drawn in the vortex penetration and expulsion regimes and phase transitions driven by temperature between vortex states are thus predicted in the zero field cooled regime before being experimentally evidenced.
}

\vspace{1cm}

PACS numbers:
\begin{itemize} 
\item 74.78.Na (Mesoscopic and nanoscale systems)
\item 74.25.Bt (Thermodynamic properties)
\item 74.25.Dw (Superconductivity phase diagrams)
\end{itemize}

\vspace{1cm}

\textit{Published in Europhysics Letters}: F.R. Ong and O. Bourgeois, \textit{Europhys. Lett.}, {\bf 79} 67003 (2007)

\vspace{0.4cm} 
doi : 10.1209/0295-5075/79/67003  

\newpage

\section{Introduction}

The effect of the topology of mesoscopic samples on basic physical properties has been demonstrated \cite{moshNature}. 
On small systems many original effects \cite{imry} appear linked to the reduced size, such as enhanced surface effects, quantum phase coherence, phonon or electrons confinement or quantization of energy. However properties of phase transitions likely to occur in nanosystems remain little explored. Heat capacity analysis is one of the most powerful tools to investigate every phase transition whatever its origin. For instance the critical temperature of a superconducting nanograin or the Curie temperature of a magnetic aggregate are hard to define through magnetization measurements since large fluctuations smear out the relevant phase transition. 

In the field of mesoscopic superconductivity, physics is dominated by the fluxoid quantization. For instance in the case of a thin doubly connected superconductor this constraint leads to physical properties that are flux periodic, the periodicity being $\Phi_0=h/2e$, the superconducting flux quantum. The Little-Parks effect \cite{little,tinkham,ongPRB06} is one of its more obvious manifestation. Oscillations with periodicity $n\times \Phi_0$ linked to the metastability of giant vortex states \cite{finkPR151} have also been reported in doubly connected mesoscopic superconductors \cite{vodoPRB03,bourgprl}. Dealing with simply connected mesoscopic superconductors the physics becomes richer and far less trivial, since fluxoid quantization do not directly leads to flux periodic properties.
A pioneering work in the late 1980's by Buisson \textit{et al.} \cite{buissonPLA90} has shown the effect of edge states of micron-sized disks on the critical temperature and on the magnetization, both exhibiting oscillations as function of magnetic field. Later in the 1990's critical field of mesoscopic superconductors of various geometries and topologies have been studied both experimentally \cite{moshNature} and theoretically \cite{benoist}. Ginzburg-Landau description is used and precise geometry of the system is taken into account through the boundary conditions of the order parameter on surfaces. In ref.~\cite{geimNature97} Geim \textit{et al.} evidence size effects through magnetization measurements carried out on single disks far from $T_c$. The shape and the features of magnetization curves strongly depend on the value of the ratio $R/\xi(T)$, where $R$ is the disk radius and $\xi(T)$ the superconducting coherence length. The order of the superconducting to normal (SN) phase transition is size dependent, and large radii disks exhibit first order phase transitions inside the superconducting region. These transitions are attributed to the entrance or exit of individual vortices. 

In this Letter we present the first heat capacity study of phase transitions between vortex states in mesoscopic simply connected superconductors. 
Such measurements have been reported only once in an unfructuous attempt \cite{lindell} to investigate the paramagnetic Meissner effect \cite{moshPRB1997,geimNature98,singhaEPL00}. Yet calorimetric investigation of such systems can bring an innovative and complementary point of view on these systems \cite{singhaEPL00,bourgprl,ongPRB06}, and recent progess in nanocalorimetry \cite{bourgprl,fonNlett05} make it now possible to measure the heat capacity of submicron-sized systems. In this work we show that the heat capacity $C_{\rm p}$ has a non trivial behaviour under an applied magnetic field $\vec{H}$. Heat capacity curves versus $T$ or $H$ are found to depend on the number of vortices threading a single disk, and the entrance or exit of a vortex causes a $C_{\rm p}$ slip from one curve to another one. This effect is particulary striking when looking at the height $\Delta C(H)$ of the $C_{\rm p}$ jump at the SN transition when sweeping the temperature under fixed $H$. Indeed $\Delta C(H)$ is strongly modulated and encounters sharp non periodic discontinuities. To interpret this behaviour we draw the full phase diagrams of these giant vortex states describing the order parameter in the disks, both in increasing and decreasing fields, and we discuss the stability of these states under different situations of magnetic field history (zero field cooled and field cooled regimes).

\section{System and apparatus}
\begin{figure}
\includegraphics[width=12cm,angle=0]{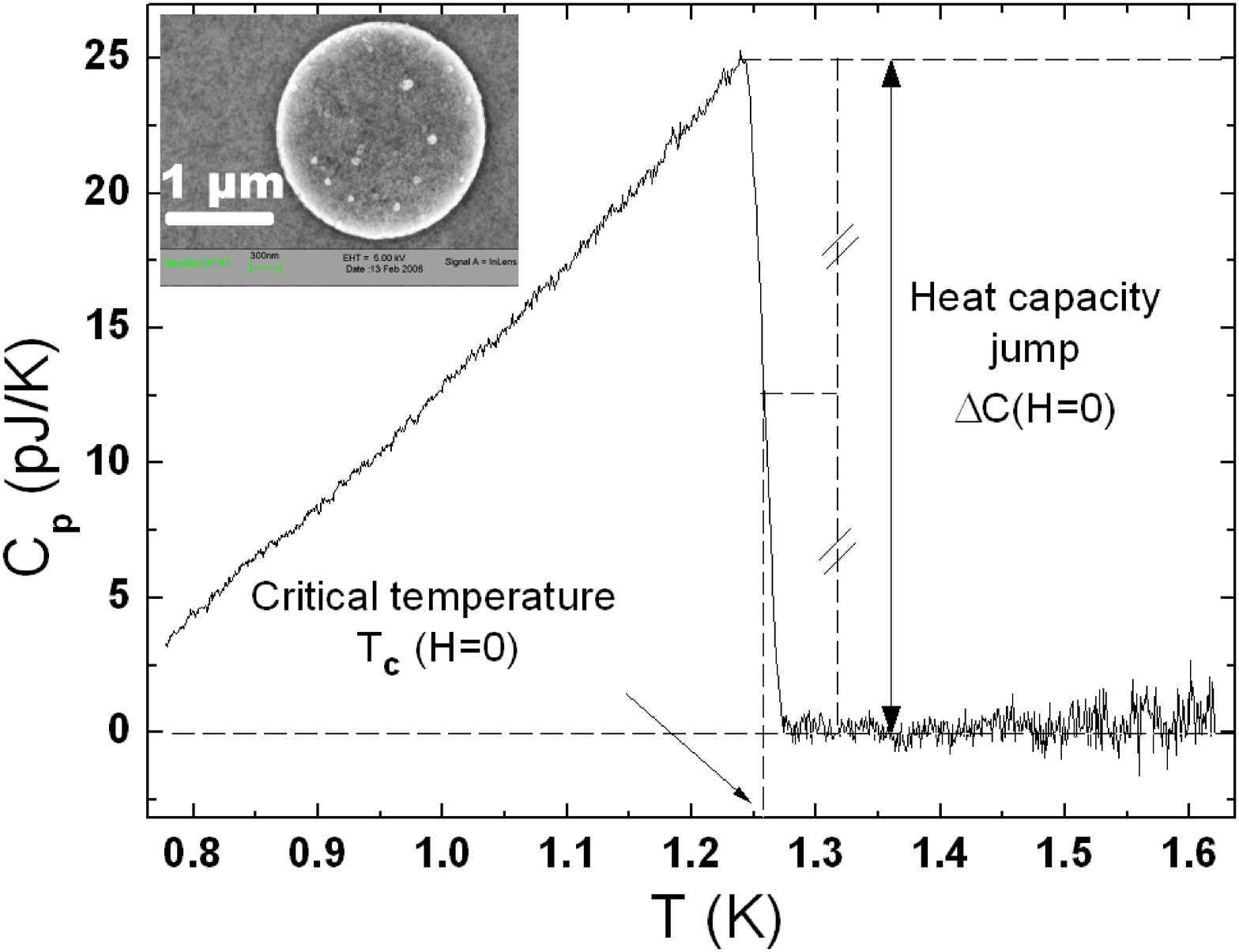}
\caption{(color online) Main plot : Heat capacity $C_{\rm p,H=0}$ of superconducting aluminium disks versus temperature under zero magnetic field. Critical temperature and height of the heat capacity jump at the superconducting to normal transition are extracted. Inset : SEM image of a single disk.}
\label{fig1}
\end{figure}

The sample studied in this work is composed of an array of $N=4.02 \times 10^5$ aluminium disks of radius $R=1.05$ $\mu$m and of thickness $e=160$ nm (see inset of fig.~\ref{fig1}), giving a total aluminium mass of $603$ ng. Despite their large number all the structures have the same geometric parameters within the measurement accuracy of the scanning electron microscope ($\approx$ 10 nm). Furthermore the separation of $2$ $\mu$m between two adjacent disks' edges ensures that the disks are non interacting and that the thermal signals are additive. These two points ensure that the measured $C_{\rm p}$ of the aluminium is $N$ times the heat capacity of a single disk, all the disks being considered identical. From an independent measure of the mean free path we estimate the Ginzburg Landau coherence length at zero temperature to be $\xi(0)\approx 0.17$ $\mu$m, and so $R/\xi(0)\approx6$ and $e/\xi(0)\approx1$, a geometry that should prevent the appearance of multivortex states \cite{schweigPRL98,kandaPRL2004}.

The $N$ mesoscopic disks are patterned by electron beam lithography on a home-made specific heat sensor and aluminum is deposited by thermal evaporation. The sensor \cite{fom97,bourgprl} is composed of a large ($4$mm$\times4$mm) and thin ($10$ $\mu$m) silicon membrane suspended by twelve silicon arms. On this membrane, a copper heater and a NbN thermometer are deposited through regular photolithography. Resistances of these thin film transducers are measured by a four point probe technique,
enabling the measurement of the total heat capacity by ac-calorimetry \cite{sullivan}. 
This method has been largely described in numerous publications \cite{bourgprl,sullivan,riou97,fom97}. The calorimetry setup is cooled down to 0.55 K using a $^3$He cryostat and is placed in the center of a large superconducting coil so that a homogenous magnetic field $\vec{H}$ can be applied perpendicular to the plane of the disks. 
We reach the quasiadiabatic conditions \cite{sullivan} for a heating current of frequency $f \approx 765$ Hz at 1 K.
The amplitude $\delta T_{\rm ac}$ of the oscillations of temperature is set in the range 1 mK to 20 mK depending on the working temperature and on the resolution needed for the measurements. The ac-calorimetry enables averaging of the measured signal: typically by averaging over 10 seconds, this apparatus allows measurements of heat capacity within 10 femto-Joule per Kelvin.

\section{Heat capacity versus temperature under fixed magnetic field}

In this section we study the influence of a constant perpendicular magnetic field on the shape of the superconducting transition (critical temperature, height, width). The sample is cooled after the field $H$ is applied (FC regime) and we measure its heat capacity versus $T$. Our aim here is to study the dependence on $H$ of the discontinuity $\Delta C$ of heat capacity at the SN transition. In a bulk geometry $\Delta C$ is monotonous with $H$ whereas in a doubly connected superconductor $\Phi_0$-periodical modulations of $\Delta C(H)$ have been reported \cite{ongPRB06}. 
To access $\Delta C$ from our raw data we have to extract the contribution $C_{\rm p,H}(T)$ of the superconducting aluminium to the total heat capacity 
(normal Al + superconducting Al + addenda) directly measured by our apparatus. This is achieved by substracting the heat capacity recorded under a high magnetic field that destroys superconductivity, as it was explained in ref.~\cite{ongPRB06}.
The heat capacity $C_{\rm p,H=0}(T)$ of the superconducting aluminium under zero magnetic field is shown on fig.~\ref{fig1}. Increasing or decreasing $T$ scans show no difference whatever the applied field.
From this curve we can extract the height $\Delta C(H)$ of the heat capacity discontinuity at the SN transition, and the critical temperature $T_c$ which is defined at the middle height of the $C_{\rm p}$ jump.

\begin{figure}
\includegraphics[width=12cm,angle=0]{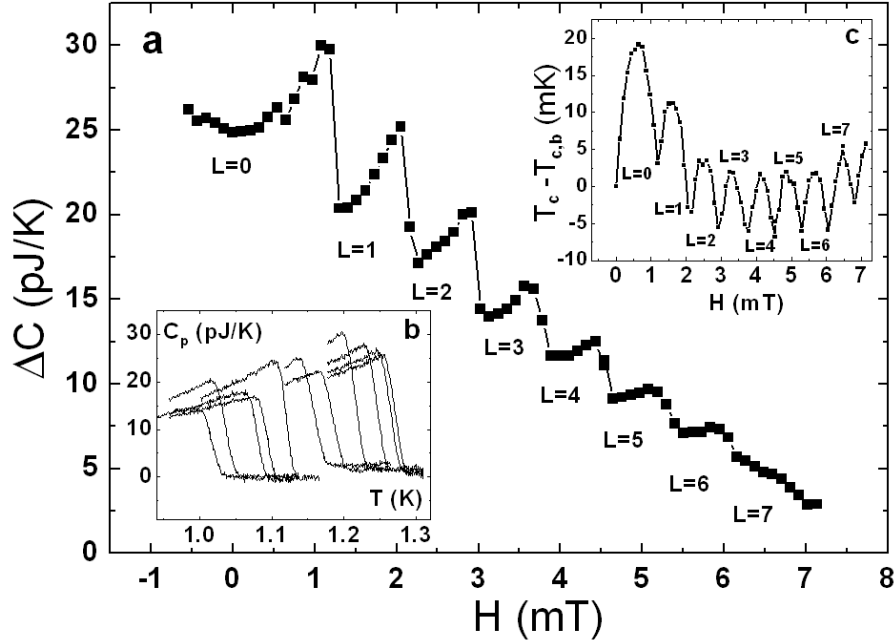}
\caption{a) Height $\Delta C(H)$ of the heat capacity discontinuity at the SN transition. b) $C_{\rm p,H}(T)$ plots under fixed magnetic fields ranging from $0$ mT (right curve) to 3.3 mT (left curve); the $i^{th}$ curve from the right has been scanned under $H=(i-1)*0.33 mT$ in the field cooled (FC) regime. c) Critical temperature $T_c(H)$ after removing a bulk-like linear trend $T_{c,\rm b}(H)$ in order to highlight the oscillating component.}
\label{fig2}
\end{figure}

We perform $C_{\rm p,H}(T)$ measurements for several fixed magnetic fields $H$ ranging from $-0.5$ to $7.2$ mT in the FC regime. Such scans are shown on fig.~\ref{fig2}.b for $H=0$ (right curve) to $3.3$ mT (left curve). As $H$ is increased we observe an original behaviour at the SN transition: as the critical temperature $T_c(H)$ regularly decreases, the heat capacity discontinuity $\Delta C(H)$ exhibits a non monotonic behaviour. $\Delta C(H)$ is measured and presented on fig.~\ref{fig2}.a: discontinuities of $\Delta C(H)$ are clearly evidenced. Heat capacity measurements on doubly connected mesoscopic superconductors (rings) have shown a modulation of $\Delta C(H)$, but no such discontinuities occured \cite{ongPRB06}. In ref~\cite{singhaEPL00} a modulated $\Delta C(H)$ was calculated for disks and no discontinuity was predicted; but the calculation was based on the linearised Ginzburg-Landau equations and the system was considered 2D. In our case, the thickness $e$ is close to the penetration length of the field ($\lambda(0)\approx70$ nm) so we cannot assume a homogenous magnetic field inside the disks. A full 3D treatment of the Maxwell equations coupled to complete Ginzburg-Landau equations is required \cite{singhaPRL97}. 

Fine measurements of the critical temperature of the disks reveal a non monotonic dependence of $T_c(H)$ which is superimposed on a bulk-like linear trend $T_{c, \rm b}(H)$. This feature is enhanced on fig.~\ref{fig2}.c where this linear background as been removed. Discontinuities of $\Delta C(H)$ and of $d T_c/d H$ occur at the same magnetic fields. These modulations appear to be non periodical, contrary to the rings' case: the pseudoperiod of the first modulation corresponds to $1.9 \Phi_0$, where $\Phi_0=h/2e$ is the superconducting flux quantum calculated through the edge of a disk. As $H$ increases the pseudoperiod decreases and tends towards a constant value of $1.5 \Phi_0$ at large $H$, which is in agreement with $T_c(H)$ measurements performed by resistance mesurements in ref.~\cite{moshNature}. This non-periodicity is due to the fact that fluxoid is quantized through a non rigid contour as it would be in a ring-like geometry \cite{bourgprl,ongPRB06}. At large magnetic fields the simply connected behaviour (constant periodicity) is recovered since the disks are composed of normal metal in the center, the superconductivity being confined on the edge of the superconductors.

Indeed from ref.~\cite{moshNature} it appears that parabola-like branches in the SN phase diagram $T_c(H)$ (fig.~\ref{fig2}.c) correspond to discrete $L$ states of angular momentum, where the so called vorticity $L$ is an integer. According to the geometry of the disks (large thickness) the authorized states minimizing the Ginzburg-Landau free energy are giant vortex states \cite{finkPR151,schweigPRL98}, whose order parameter has the form $\Psi(\vec{r})=f(z,r)exp(iL\theta)=f(r)exp(iL\theta)$ since the thickness $e$ is smaller than $\xi(T) \approx 1$ $\mu$m at $1$ K. $L$ is the number of single vortices threading a single disk and corresponds to the number of fluxoid quanta trapped in. A giant vortex state is a state of edge superconductivity. Indeed $L$ vortices share the same core so superconductivity is located near the boundary of the disks. Such a state is not stable in a bulk geometry where vortex-vortex repulsion is not balanced by the vortex-edge repulsion that has to be taken into account in a mesoscopic superconductor. Thus discontinuities of $\Delta C(H)$ on fig.~\ref{fig2}.b correspond to jumps between successive $L$ states : as the sample is cooled below its critical temperature under an applied magnetic field $H$, the disks transit from normal state to a superconducting state with vorticity $L$ only depending on $H$, and remain in this $L$ state as temperature is decreased. This issue will be made clearer in the following. Then as $C_{\rm p,H}(T)$ is scanned by increasing $T$ we eventually measure $C_{\rm p,H,L}(T)$ at a fixed $H$ \textit{and }$L$.

The strong modulations of $\Delta C(H)$ on fig.~\ref{fig2}.a show that heat capacity can be modulated by an external parameter, here the magnetic field. A consequence is that heat capacity looses its additive property in a mesoscopic superconductor, and more widely in a mesoscopic system. Indeed if one calculates the specific heat of a given mesoscopic superconductor by dividing its heat capacity by its mass, one would obtain a different result depending on the way the system is structured : simply or doubly connected, square or ring etc. Thus the intensive \textit{specific heat} has no sense at the mesoscale. This original property of small systems has been proposed in ref.~\cite{ongPRB06} but the large error bars due to the small measured mass made this point controversial. In the present work non additivity of the heat capacity in small systems is unambiguoulsy evidenced. Furthermore the influence of the topology on the heat capacity of a mesoscopic superconductor appears clearly when the shapes of $\Delta C(H)$ on disks (fig.~\ref{fig2}.a) and on rings \cite{ongPRB06} are compared.

\begin{figure}
\includegraphics[width=12cm,angle=0]{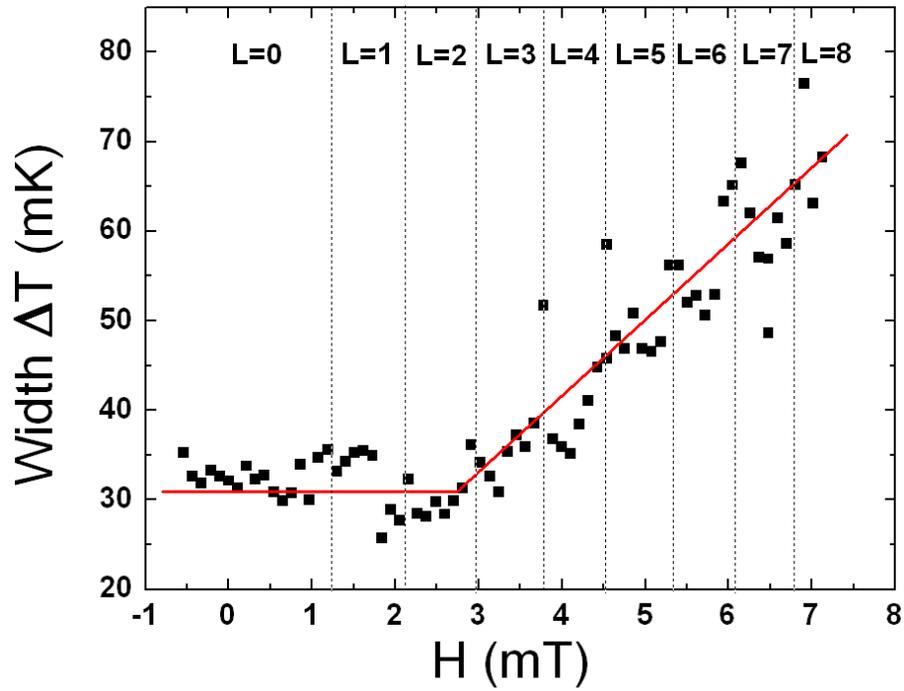}
\caption{Width of the SN transition on $C_{\rm p,H}(T)$ plots around $T_c(H)$. $L$ is the $H$-dependent vorticity describing the state of the disks. In the region $L=1$ we observe a crossover from a bulk-like behaviour to a mesoscopic behaviour. }
\label{fig2bis}
\end{figure}

Another mesoscopic effect lies in the temperature broadening $\Delta T$ of the SN transitions on the $C_{\rm p,H}(T)$ plots. $\Delta T$ is plotted versus $H$ on fig.~\ref{fig2bis}. For a small applied field ($H<3$ mT) the width $\Delta T \approx 30$ mK is field independent and has two main origins: the measurement process itself which averages the heat capacity over $\approx 20$ mK, and the slight dispersion in the geometrical parameters of the disks. As the field is increased beyond $3$ mT, $\Delta T$ starts to grow: there is a crossover between a situation where $C_{\rm p,H}(T)$ has a discontinuity at $T_c$ and another one where the SN transition is smeared out. From ref.~\cite{singhaEPL00} we interpret this as a crossover from a bulk-like behaviour to a mesoscopic behaviour. Indeed at low field surface effects can be neglected since edges are taken into account only through a boundary condition that becomes irrelevant in the absence of field \cite{singhaEPL00}; thus we expect a sharp discontinuity characteristic of second order phase transitions in bulk materials. At higher fields the constraint of vanishing supercurrents normal to surfaces becomes relevant and so a broadening of the transition appears which is characteristic of finite samples. From fig.~\ref{fig2bis} we state that this crossover occurs when two vortices have entered the disks (state $L=2$); in zero field the mesoscopic nature of the disks cannot be evidenced by thermal measurements.

\section{Heat capacity versus magnetic field under fixed temperature}

\begin{figure}
\includegraphics[width=12cm,angle=0]{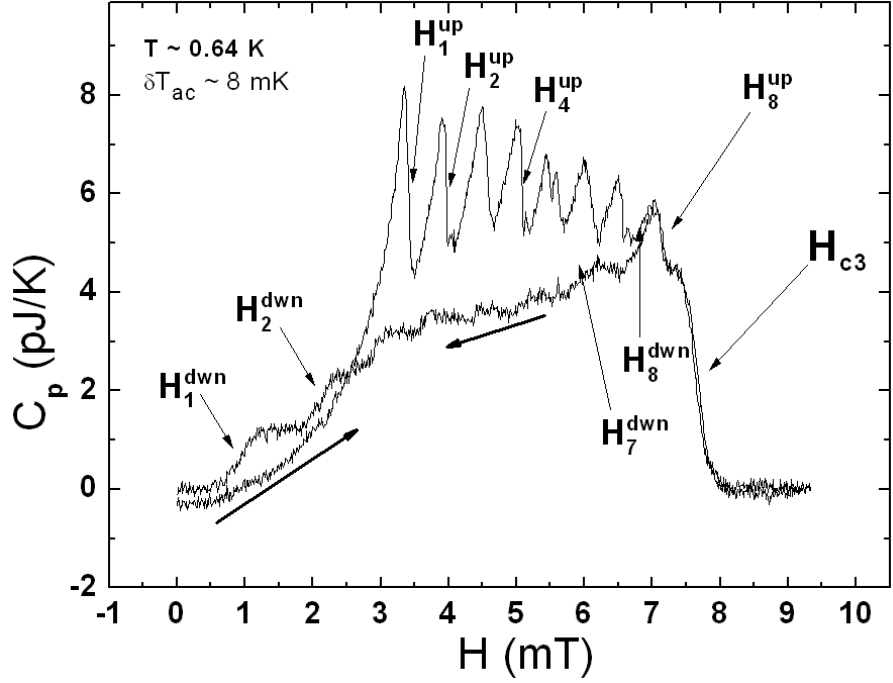}
\caption{Heat capacity $C_{\rm p, T}(H)$ versus magnetic field at fixed temperature $T=$0.64 K, in increasing and decreasing fields. The data show strong hysteresis. $H_L^{\rm up}$ is the penetration field of the $L^{\rm th}$ vortex and $H_L^{\rm dwn}$ its expulsion field. Both are measured at middle height of the $C_{\rm p}$ jump. $H_{\rm c3}$ is the critical field beyond which superconductivity is suppressed.}
\label{fig3}
\end{figure}

In order to highlight the possibility of modulating the heat capacity of the mesoscopic disks with $H$, we perform the counterpart of previously described measurements: we fix the temperature before scanning the heat capacity $C_{\rm p, T}(H)$ versus magnetic field. An example is presented in fig.~\ref{fig3} for $T=0.64$ K. We present $C_{\rm p, T}(H)$ in increasing and decreasing magnetic fields. Starting from zero field, the disks first remain in the Meissner state ($L=0$) until $H\approx 3.3$ mT where $C_{\rm p}$ suddenly shifts to a lower value: this $C_{\rm p}$ jump is the signature of the first order phase transition $L=0\rightarrow L=1$ : at this point a vortex penetrates each disk. Then as the field is increased one can observe a cascade of successive phase transitions $L\rightarrow L+1$, until the SN second order transition occuring at $H_{\rm c3}$ (here we use $H_{\rm c3}$ instead of $H_{\rm c2}$ to express the critical field because the giant vortex states describing the order parameter are surface superconductivity states). The change of symmetry related to these phase transitions concerns the order parameter $\Psi(\vec{r})=f(r)exp(iL\theta)$ whose phase is invariant under a rotation of angle $2\pi/L$. By adding a quantum of angular momentum the order of symmetry shifts from $L$ to $L+1$. These phase transitions are not periodic with $H$ for the same reason as described in the previous section (absence of rigid contour to quantize fluxoid). Now looking at the $C_{\rm p, T}(H)$ plot in decreasing field, it appears that $C_{\rm p, T}(H)$ is strongly hysteretic, which is consistent with magnetization measurements of ref.~\cite{geimNature97}. $C_{\rm p}$ jumps are also visible although less pronounced than in increasing $H$, and are signatures of phase transitions of type $L\rightarrow L-1$: at each jump the disks expell a vortex. Phase transitions of type $L \rightarrow L \pm 1$ are of first order \cite{geimNature97} and a non vanishing latent heat proportional to $(S_{\rm L}-S_{\rm L\pm 1})$ is involved.

From the data of fig.~\ref{fig3} we can define the penetration (resp. expulsion) field $H_L^{\rm up}$ (resp. $H_L^{\rm dwn}$) of the $L^{\rm th}$ vortex. Both are measured at middle height of the $C_{\rm p}$ jump. We notice that $H_L^{\rm up}>H_L^{\rm dwn}$. The origin of this hysteresis is related to the occurence of metastability. According to ref.~\cite{singhaPRB99} this metastability is caused by the surfacic Bean-Livingston barrier \cite{bean}: supercurrents around a vortex circulate near the edges in the opposite direction of the currents which screen the field, leading to a repulsive force between edge and vortices. This interaction prevents the nucleation of a vortex on an edge although the penetration of such a vortex could lower the free energy. However this surfacic barrier is suppressed in increasing magnetic field due to pinning on surface rugosity, which is not the case in decreasing magnetic field \cite{singhaPRB99}. Thus on fig.~\ref{fig3} the heat capacity appears to be the one of the fundamental state when $H$ is swept up, whereas metastability occurs as $H$ is swept down. It is noteworthy that the width of the $C_{\rm p}$ jumps in increasing field is small ($\approx 0.1$ mT) compared to the pseudo period ($\approx 0.55$ mT) of $C_{\rm p, T}(H)$. This means that all the disks transit at a well defined magnetic field $H_L^{\rm up}$ corresponding to the field beyond which the state $L+1$ becomes thermodynamically more stable than state $L$, without considering the individual disorder of a disk. On the other hand phase transitions $L\rightarrow L-1$ in decreasing field occur at magnetic fields that are not so well defined as in increasing field. The reason for this is that the field at which the metastable state $L$ becomes unfavourable compared to state $L-1$ depends on the surface defects of a given disk. Thus the distribution of microcsopic disorder among disks leads to a broadening of the $L\rightarrow L-1$ transition width.

\section{Giant vortex states phase diagrams}

\begin{figure}
\includegraphics[width=12cm,angle=0]{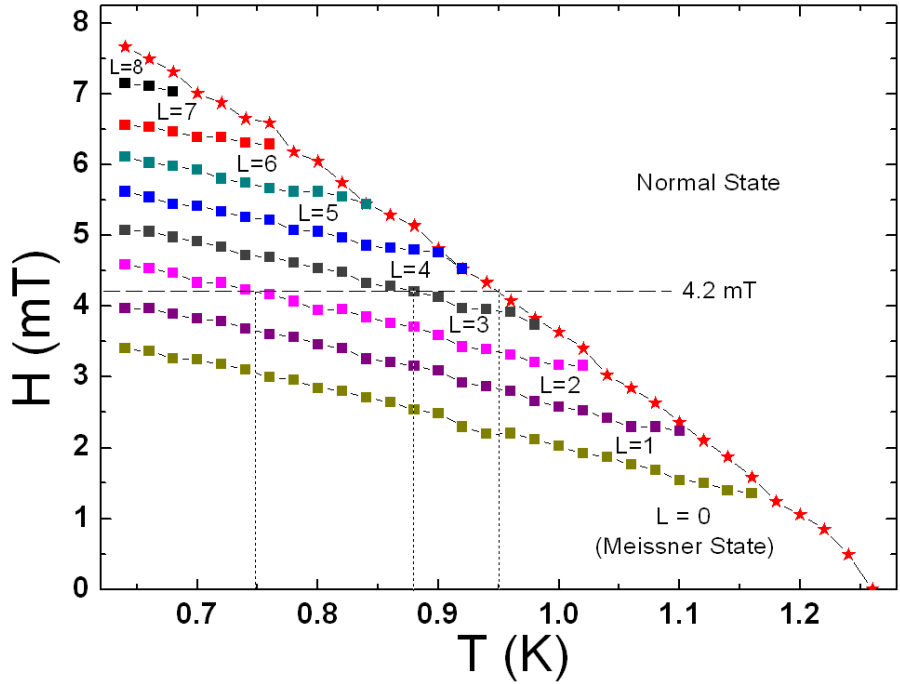}
\caption{(color online) Phase diagram of the disks in increasing magnetic field. The stars represents the SN transition line. The squares are the penetration fields $H_L^{\rm up}$ versus $T$. The dashed line corresponds to a magnetic field of 4.2 mT and the dotted lines localize its intersections with the different vorticity lines.}
\label{fig4}
\end{figure}
\begin{figure}
\includegraphics[width=12cm,angle=0]{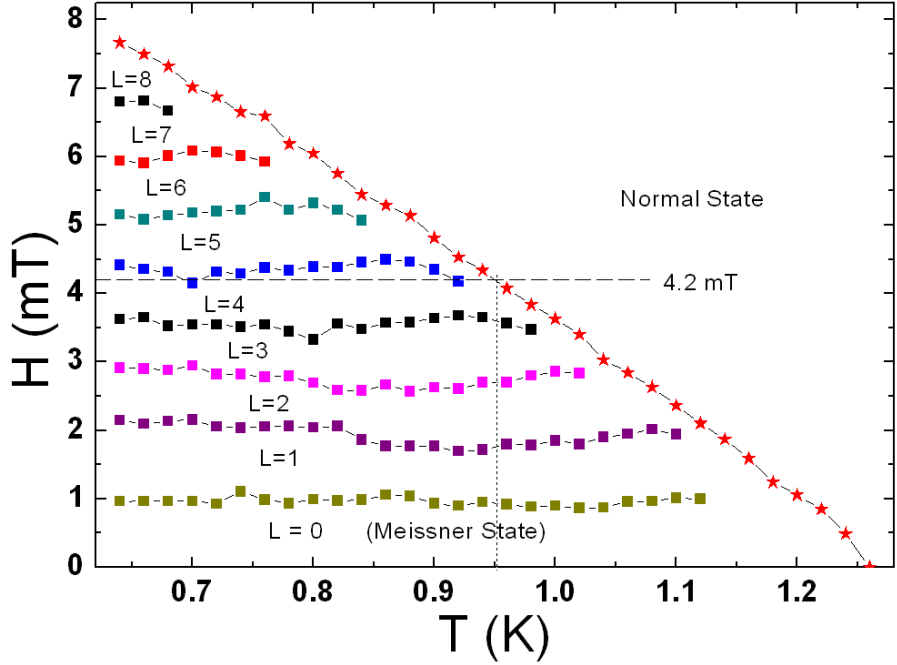}
\caption{(color online) Phase diagram of the disks in decreasing magnetic field. The SN transition line (stars) is the same as in fig.~\ref{fig4}. The squares are the expulsion fields $H_L^{\rm dwn}$ versus $T$. Near the critical temperature transitions $L\rightarrow L-1$ were hard to localize because of their small height compared to the noise, and so $H_L^{\rm dwn}(T)$ was not measurable near the SN line. The dashed line corresponds to a magnetic field of 4.2 mT and the dotted line localizes its intersection with the normal state line.}
\label{fig5}
\end{figure}

By repeating $C_{\rm p, H}(T)$ scans at many fixed temperatures ranging from $0.64$ K to $T_c=1.26$ K, 
we observe that as $T$ increases the number of successive transitions lowers since $H_{\rm c3}$ decreases. Another effect is that the penetration fields $H_L^{\rm up}$ shifts to lower fields as $T$ is increased. To hightlight this effect we plot the dependence of $H_L^{\rm up}$ versus temperature $T$ for the 9 first $L$ states accessible in our experiment: the corresponding phase diagram is presented on fig.~\ref{fig4}. In the same way we present the temperature dependence of expulsion fields $H_L^{\rm dwn}$ on fig.~\ref{fig5}. Fig.~\ref{fig5} is much more noisy than fig.~\ref{fig4} since the widths (resp. the heights) of $L\rightarrow L-1$ transitions are larger (resp. smaller) than those of $L\rightarrow L+1$ transitions: both effects make it hard to localize precisely expulsion fields.

Despite this uncertainty when measuring $H_L^{\rm dwn}$ it appears clearly that $H_L^{\rm up}(T)$ and $H_L^{\rm dwn}(T)$ do not behave the same way: $H_L^{\rm up}(T)$ decreases as $T$ increases whereas $H_L^{\rm dwn}(T)$ is almost temperature independent. These observations are in good aggreement with ref.~\cite{baelusPRB05} and also are complementary to that work. Indeed, Baelus \textit{et al.} calculated penetration and expulsion fields of mesoscopic disks (with a slightly different geometry) and measured them through a magnetization study at low temperature. Since magnetization is weak close to $T_c$ they report measurements only from $0.1$ to $0.5$ K. Within this interval and for the 8 first $L\rightarrow L \pm 1$ transitions they also found a penetration field that decreases with $T$ and a temperature independent expulsion field in the case of Giant Vortex States. The results we present in this paper extend this property to all the phase diagram including the close-to-$T_c$ area. Furthermore, according to ref. \cite{baelusPRB05}, the monotonous behaviour of the $H_L^{\rm dwn}(T)$ lines confirms our assumption that our disks can only host giant vortex states, since the expulsion fields of multivortex states would increase with $T$.

\begin{figure}
\includegraphics[width=12cm,angle=0]{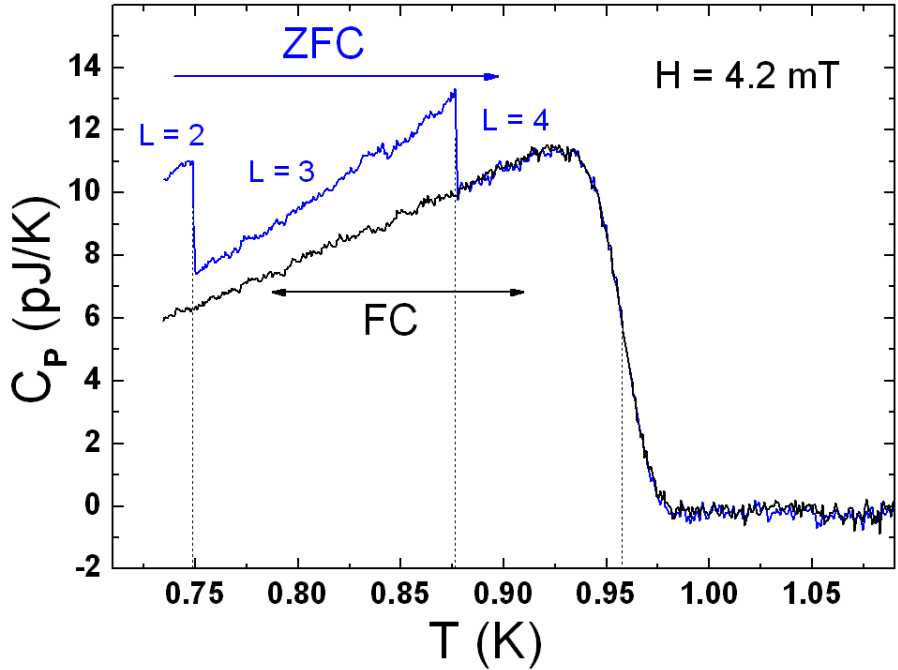}
\caption{(color online) $C_{\rm p,H}(T)$ plots under fixed $H=4.2$ mT in FC (black curve) and ZFC regimes (blue curve).}
\label{fig6}
\end{figure}

The different behaviours of penetration and expulsion fields versus $T$ can be exploited to predict phase transitions of type $L\rightarrow L+1$ driven by temperature (successive jumps in $C_{\rm p,H}(T)$) instead of $H$ as in the last section \cite{geimNature98}. Indeed up to now we have presented $C_{\rm p,H}(T)$ plots that were scanned in the field cooled (FC) regime. In that case as the system is cooled from the normal state to the superconducting state it then remains in the same $L$ state as $T$ is swept up or down, leading to the observation of the only regular SN phase transition. In the zero field cooled (ZFC) regime the situation is quite different. To illustrate this issue we present on fig.~\ref{fig6} $C_{\rm p,H}(T)$ plots scanned under a fixed $H=4.2$ mT both in FC and ZFC regimes and we interpret them using phase diagrams shown on figs.~\ref{fig4} and \ref{fig5}. In the FC regime the system evolves along the dashed horizontal line $H=4.2$ mT of fig.~\ref{fig5} since the magnetic field is already plugged: once the system has transited into the superconducting state, it remains in the same $L=4$ state, whatever the temperature. Thus when increasing $T$ to scan $C_{\rm p,H}(T)$, the only expected phase transition occur at the SN boundary. This is what is observed on the lower curve of fig.~\ref{fig6}. On the other hand in the ZFC regime we plug the magnetic field once the system is cooled down. Then when increasing $T$ the system has to evolve along the dashed horizontal line $H=4.2$ mT of fig.~\ref{fig4}: this line crosses several regions of the phase diagram inside the superconducting area leading to several phase transitions ($L=2\rightarrow 3$ and $L=3\rightarrow 4$) before the SN transition ($L=4\rightarrow N$). Fig.~\ref{fig6} shows the experimental heat capacity signatures of these phase transitions. Such thermal signatures of vortex entrances in mesoscopic superconductors driven by temperature at constant applied magnetic field is reported here for the first time.

\section{Conclusion}

We have studied the heat capacity dependence versus $H$ and $T$ of an array of independent superconducting mesoscopic disks whose size allows the presence of giant vortex states characterized by their angular momentum $L$. Looking at the SN transition on $C_{\rm p, H}(T)$ curves we observe  a crossover from a bulk-like to a mesoscopic behaviour: the mesoscopic character arises for $L \geq 2$ since surfaces play no role in the $L=0$ state in agreement with predictions of ref.~\cite{singhaEPL00}. Another mesoscopic signature in the thermal signal is the strong modulation of $\Delta C(H)$ at the SN transition. As a consequence the concept of specific heat fails to describe mesoscopic superconductors since it becomes possible to modulate their heat capacity with an external parameter (here $H$) in a way that depends on their topology, geometry and size. Playing with the magnetic history of the sample and its hysteretical properties, we are able to draw either the $H$ or the $T$ dependence of the heat capacity $C_{\rm L}(T,H)$ at fixed vorticity $L$ in order to establish a complete phase diagram in temperature and magnetic field.

 
\section*{Acknowledgments}

We would like to thank E. Andr\'e, P. Lachkar, J-L. Garden, C. Lemonias, B. Fernandez, T. Crozes for technical support, J. Chaussy, P. Brosse-Maron, T. Fournier, Ph. Gandit and J. Richard for fruitfull discussions and help. We thank the Région Rhône-Alpes for the PhD grant of F.R. Ong and the IPMC (Institut de Physique de la Mati\`ere Condens\'ee) of Grenoble for financing a part of this project.

\newpage

\end{document}